\documentclass[twocolumn]{aastex701}

\usepackage{xcolor}
\usepackage{amsmath}
\usepackage{booktabs, tabularx, makecell,array,threeparttable}
\usepackage{hyperref}
\usepackage{rotating}

\begin{document}
\title{Spectral Energy Distribution and Polarimetric Properties of Dust in the Interstellar Comet 3I/ATLAS}

\author[orcid=0009-0006-1028-1653]{Seungwon Choi}
\affil{Department of Physics and Astronomy, Seoul National 
University, 1 Gwanak-ro, Gwanak-gu, Seoul 08826, Korea}
\affil{SNU Astronomy Research Center, Department of Physics and Astronomy, Seoul National University, 1 Gwanak-ro, Gwanak-gu, Seoul 08826, Republic of Korea}
\email[show]{cygalbireo@snu.ac.kr}

\author[orcid=0000-0002-7332-2479]{Masateru Ishiguro}
\affiliation{Department of Physics and Astronomy, Seoul National 
University, 1 Gwanak-ro, Gwanak-gu, Seoul 08826, Korea}
\affiliation{SNU Astronomy Research Center, Department of Physics and Astronomy, Seoul National University, 1 Gwanak-ro, Gwanak-gu, Seoul 08826, Republic of Korea}
\email[show]{ishiguro@snu.ac.kr}

\author[orcid=0000-0001-9067-7653]{Hiroyuki Naito}
\affiliation{Nayoro Observatory, 157-1 Nisshin, Nayoro, Hokkaido, 096-0066, Japan}
\email{}

\author[orcid=0000-0002-2618-1124]{Yoonsoo P. Bach}
\affiliation{Korea Astronomy and Space Science Institute (KASI), 776 Daedeok-daero, Yuseong-gu, Daejeon 34055, Republic of Korea}
\email{}

\author[orcid=0000-0002-3291-4056]{Jooyeon Geem}
\affiliation{National Astronomical Observatory of Japan, Mitaka 181-8588, Japan}
\email{}

\author[orcid=0000-0002-8244-4603]{Bumhoo Lim}
\affiliation{Department of Physics and Astronomy, Seoul National 
University, 1 Gwanak-ro, Gwanak-gu, Seoul 08826, Korea}
\affiliation{SNU Astronomy Research Center, Department of Physics and Astronomy, Seoul National University, 1 Gwanak-ro, Gwanak-gu, Seoul
08826, Republic of Korea}
\email{}

\author[orcid=0009-0001-3706-5671]{Moonkuk Hur} %
\affiliation{Department of Physics and Astronomy, Seoul National 
University, 1 Gwanak-ro, Gwanak-gu, Seoul 08826, Korea}
\affiliation{SNU Astronomy Research Center, Department of Physics and Astronomy, Seoul National University, 1 Gwanak-ro, Gwanak-gu, Seoul
08826, Republic of Korea}
\email{}

\author[orcid=0009-0002-4850-9167]{Sihu Ahn} %
\affiliation{Department of Physics and Astronomy, Seoul National 
University, 1 Gwanak-ro, Gwanak-gu, Seoul 08826, Korea}
\affiliation{SNU Astronomy Research Center, Department of Physics and Astronomy, Seoul National University, 1 Gwanak-ro, Gwanak-gu, Seoul
08826, Republic of Korea}
\email{}

\author[orcid=0000-0002-0460-7550]{Sunho Jin}
\affiliation{Korea Astronomy and Space Science Institute (KASI), 776 Daedeok-daero, Yuseong-gu, Daejeon 34055, Republic of Korea}
\email{}

\author[orcid=0000-0003-0693-1618]{Hisayuki Kubota}
\affiliation{Department of Physics and Earth Sciences, Faculty of Science, University of the Ryukyus}
\email{}

\author[orcid=0000-0002-7084-0860]{Seiko Takagi}
\affiliation{Department of Earth and Planetary Sciences, Faculty of Science, Hokkaido University, Kita-ku, Sapporo, Hokkaido 060-0810,
Japan}
\email{}

\author[orcid=0000-0001-6184-9671]{Tatsuharu Ono}
\affiliation{Department of Earth and Planetary Sciences, Faculty of Science, Hokkaido University, Kita-ku, Sapporo, Hokkaido 060-0810,
Japan}
\email{}

\author[0000-0003-2831-0513]{Max Mahlke}
\affiliation{Telespazio UK S.L. for the European Space Agency (ESA), European Space Astronomy Centre (ESAC), Camino Bajo del Castillo s/n, 28692 Villanueva de la Cañada, Madrid, Spain}
\email{}

\author[0000-0002-9548-1526]{Carey	Lisse}
\affiliation{Johns Hopkins University Applied Physics Laboratory, 11100 Johns Hopkins Rd, Laurel, MD 20723, USA}
\email{}

\author[orcid=0000-0003-1726-6158]{Tomohiko Sekiguchi}
\affiliation{Asahikawa Campus, Hokkaido University of Education, Hokumon 9, Asahikawa, Hokkaido 070-8621,Japan}
\email{}

\author[orcid=0000-0002-3656-4081]{Makoto Watanabe}
\affiliation{Department of Physics, Okayama University of Science, 1-1 Ridai-cho, Kita-ku, Okayama, Okayama 700-0005, Japan}
\email{}

\newcommand{\Rc}{\ensuremath{R_\mathrm{C}}}
\newcommand{\Ic}{\ensuremath{I_\mathrm{C}}}
\newcommand{\Ks}{\ensuremath{K_\mathrm{s}}}

\begin{abstract}
The interstellar comet 3I/ATLAS exhibits unusual polarimetric properties
compared with Solar System comets, while constraints on its fundamental physical properties, such as dust albedo and grain-size distribution, remain limited.
We obtained new optical polarimetry near opposition with the Pirka
Telescope, combined it with previously reported polarimetric measurements,
and constructed a visible-to-mid-infrared spectral energy distribution
(SED) from HST, VLT, SPHEREx, and JWST observations.
The SED indicates that the coma is dominated by dark, absorbing dust,
with an effective $V$-band albedo of $p_{\rm eff}\sim0.04$. 
Modeling the SED with a power-law grain-size distribution yields a size index of $q=3.0$--$3.3$.
The mid-infrared spectral structure is better reproduced by adding a
minor amorphous-silicate component.
The combined polarimetric data further favor a highly porous scattering
structure.
Despite its unusual polarimetric behavior, the dust of 3I/ATLAS has an albedo and grain-size distribution within the range found for Solar System comets.
\end{abstract}


\section{Introduction}

The physical properties of cometary dust have been investigated using complementary constraints from its spectral energy distribution (SED) and polarimetric behavior \citep{2024come.book..577E}. From visible to mid-infrared wavelengths, the dust continuum changes from predominantly scattered sunlight to thermal emission, with both components contributing at intermediate wavelengths. Solid-state features, particularly the prominent silicate emission around $10~\mu\mathrm{m}$, provide additional constraints on dust composition and grain properties
\citep{1994ApJ...425..274H}.
Polarimetry provides complementary information on the light-scattering properties of dust, including particle size, refractive index, porosity, and aggregate structure \citep{2006A&A...449.1243K}. Most Solar System comets (SSCs) exhibit broadly similar polarization phase curves (PPCs) at low phase angles \citep{2025ApJ...983L..19L}, although a dichotomy at larger phase angles has been debated \citep{1996A&A...313..327L}.

The discovery of interstellar objects (ISOs) provides an opportunity to extend such studies to primitive materials formed in planetary systems around other stars \citep{2023ARA&A..61..197J}. The first identified ISO, 1I/`Oumuamua, showed no detectable dust coma \citep{2017Natur.552..378M}, whereas 2I/Borisov exhibited a prominent cometary dust coma and tail \citep{2020NatAs...4...53G}. The third known ISO, 3I/ATLAS (C/2025 N1), was discovered on 2025 July 1 at a heliocentric distance of approximately $4.5$ au \citep{2025MPEC....N...12D}.

Observations of 3I/ATLAS have revealed several distinctive properties of its dust and gas coma. Polarimetry showed an exceptionally deep negative-polarization branch compared with SSCs \citep{2025ApJ...992L..29G}, and subsequent observations across perihelion showed that this unusual PPC persisted, with a wavelength dependence broadly consistent with that of SSCs, showing a polarimetric peak at $\sim1.5$--$2.0~\micron$ \citep{2026ApJ..1000L..29C}. \citet{2026arXiv260904932G} interpreted the polarimetric behavior in terms of porous, bright dust aggregates, whereas \citet{2025RNAAS...9..257Z} suggested a substantial contribution from pure water-ice particles.
Optical observations revealed a red dust continuum and the subsequent emergence of CN emission \citep{2025A&A...700L..10A,2026ApJ..1000...53P}. Near-infrared observations revealed absorption by solid H$_2$O in the coma (together with gas emission from CO$_2$, CO, and H$_2$O) 
\citep{2026ApJ..1000L..52L,2025ApJ...991L..43C},
while mid-infrared spectroscopy revealed a prominent silicate emission feature interpreted as being dominated by amorphous silicates \citep{2026arXiv260627535B}.

Previous studies have investigated the dust properties of 3I/ATLAS using various observational and modeling approaches. However, constraints on fundamental dust properties such as the grain-size distribution and albedo remain limited. In particular, the dust albedo of 3I/ATLAS has generally been assumed rather than measured \citep[e.g.,][]{2026ApJ...999L..37H}. Complementary constraints can be obtained from the spectral energy distribution (SED) by considering the energy balance between incident solar radiation and the scattered and thermally emitted radiation from the dust grains.
In this work, we constructed a visible-to-mid-infrared SED from HST, VLT, SPHEREx, and JWST observations to constrain the dust albedo, grain-size distribution, and composition. We also obtained new optical polarimetry with the Pirka Telescope at a phase angle as small as 0.70\degr, probing the polarimetric behavior of 3I/ATLAS near opposition, and combined these data with previously reported polarimetric measurements. Together, the SED and polarimetric constraints allow us to investigate the physical properties of dust in the coma of 3I/ATLAS and compare them with those of Solar System comets.


\section{OBSERVATIONS AND ARCHIVAL DATA ANALYSIS}

We obtained optical imaging polarimetry of 3I/ATLAS and analyzed archival observations to construct its visible-to-mid-infrared spectral energy distribution (SED). We describe these two datasets and their analysis below.

\subsection{Polarimetric Observations with MSI}

We conducted optical imaging polarimetry of 3I/ATLAS with the Multi-Spectral Imager (MSI) mounted on the 1.6 m Pirka Telescope at Nayoro Observatory, Hokkaido University. MSI is a beam-splitting polarimeter equipped with a rotatable half-wave plate and a Wollaston prism \citep{2012SPIE.8446E..2OW}. The observations were obtained in the Johnson--Cousins \Rc~and \Ic~bands. The effective field of view in the imaging-polarimetric mode is $3.3\times0.7$ arcmin, with a pixel scale of $0.39\arcsec$ pixel$^{-1}$. We observed 3I/ATLAS on seven nights, focusing on the near-opposition phase-angle range of $\alpha=0.7^\circ$--$3.0^\circ$, and obtained six nightly averaged polarimetric measurements in each band. The observing circumstances are summarized in Table~\ref{tab:observation_log}. The data reduction and polarimetric analysis followed the procedures described by \citet{2022MNRAS.509.4128I}. The calibration data used in the analysis and the nightly averaged polarimetric measurements are provided in Appendix~\ref{s:appendixA}.

\begin{deluxetable*}{@{\extracolsep{\fill}}cccccccc@{}}[t]
\tabletypesize{\small}
\tablewidth{\textwidth}
\tablecolumns{8}

\tablecaption{Observation log.\label{tab:observation_log}}

\tablehead{
\colhead{UT Date} &
\colhead{Filter} &
\colhead{Exptime} &
\colhead{$N$} &
\colhead{$r_{\rm H}$} &
\colhead{$\Delta$} &
\colhead{$\alpha$} &
\colhead{Airmass} \\
\colhead{} &
\colhead{} &
\colhead{[s]} &
\colhead{} &
\colhead{[au]} &
\colhead{[au]} &
\colhead{[\degr]} &
\colhead{} \\
\colhead{(1)} &
\colhead{(2)} &
\colhead{(3)} &
\colhead{(4)} &
\colhead{(5)} &
\colhead{(6)} &
\colhead{(7)} &
\colhead{(8)}
}

\startdata
2026-01-17.60 & \Ic & 60.0, 90.0, 120.0 & 3, 12, 1 & 3.191 & 2.215 & 2.913 & 1.155 \\
2026-01-17.63 & \Rc & 60.0, 90.0, 120.0 & 3, 12, 1 & 3.192 & 2.215 & 2.945 & 1.132 \\
2026-01-18.69 & \Ic & 90.0 & 12 & 3.225 & 2.246 & 2.303 & 1.183 \\
2026-01-18.72 & \Rc & 90.0 & 14 & 3.226 & 2.247 & 2.325 & 1.292 \\
2026-01-20.47 & \Rc & 90.0 & 6 & 3.281 & 2.299 & 1.351 & 1.653 \\
2026-01-20.50 & \Ic & 90.0 & 6 & 3.283 & 2.300 & 1.339 & 1.412 \\
2026-01-22.42 & \Rc & 90.0, 120.0 & 8, 3 & 3.345 & 2.360 & 0.701 & 2.192 \\
2026-01-22.43 & \Ic & 90.0 & 8 & 3.345 & 2.360 & 0.700 & 1.990 \\
2026-01-23.40 & \Ic & 90.0 & 5 & 3.377 & 2.392 & 0.826 & 2.506 \\
2026-01-24.65 & \Ic & 90.0 & 3 & 3.417 & 2.434 & 1.280 & 1.162 \\
2026-01-24.70 & \Rc & 90.0 & 3 & 3.418 & 2.435 & 1.309 & 1.299 \\
2026-01-25.65 & \Rc & 90.0 & 12 & 3.449 & 2.467 & 1.720 & 1.163 \\
\enddata
\tablenotetext{}{
\textbf{Note.—}
All quantities are evaluated at the midpoint of the observation period.
(1) UT date.
(2) Filter used in each observation.
(3) Individual exposure time.
(4) Number of individual exposures corresponding, in order, to the
exposure times listed in Column (3).
(5) Heliocentric distance.
(6) observer--object distance.
(7) Phase angle.
(8) Airmass.
The ephemerides were obtained from the JPL Horizons system
(\citealt{JPLHorizons2025}). 
}

\end{deluxetable*}

\subsection{SED Generation from Archival Data}

Optical to mid-infrared observations of 3I/ATLAS were obtained with several facilities around 2025 December. Because the observations were obtained at different epochs and with different apertures, we corrected for differences in observing geometry and applied appropriate aperture scaling to construct a single visible-to-mid-infrared SED. Details of the observational datasets used to construct the SED are provided in Appendix~\ref{s:appendixB}.

For the optical range, we used HST/WFC3 F350LP imaging obtained on 2025 December 12 and the publicly available flux-calibrated, nightly stacked VLT/ESPRESSO fiber-A spectrum obtained on December 21. The HST dataset is available from MAST at \dataset[doi:10.17909/r41g-5d23] {https://doi.org/10.17909/r41g-5d23}. 
Aperture photometry of the HST image was performed using a circular aperture of 0.496 arcsec (652 km) radius, chosen to match the aperture used for the JWST/MIRI spectrum. The sky background was estimated from circular annulus with inner radius of 40 arcsec and outer radius of 45 arcsec. The resulting HST photometry was used as the absolute optical flux reference for constructing the SED.

For the near-infrared range, we used SPHEREx imaging-spectroscopy data obtained between December 8 and 15. For the mid-infrared range, we retrieved the JWST/MIRI Stage~3 spectral cubes obtained on December 15 under DDT Program 9442. The JWST dataset is available from MAST at \dataset[doi:10.17909/cpmw-zn69] {https://doi.org/10.17909/cpmw-zn69}. 
Source and off-target background spectra were extracted using circular apertures of $0.5\arcsec$ radius, corresponding to a projected photocentric radius of approximately 650 km. Thus, the HST photometry and JWST/MIRI spectrum were extracted using matched projected apertures.

Because the ESPRESSO and SPHEREx observations were obtained with different apertures, their absolute flux levels were adjusted by multiplicative scaling factors. The ESPRESSO spectrum was scaled to match the aperture-matched HST photometry, and the SPHEREx spectrum was then scaled to match the ESPRESSO continuum over their overlapping wavelength range at $0.75$--$0.79~\mu\mathrm{m}$. The MIRI spectral segments were combined using the Channel~3 SHORT segment as the reference. The resulting dataset provides the visible-to-mid-infrared SED used in the following analysis.


\section{Results}

We first present the SED analysis and derive the dust albedo, grain-size distribution, and constraints on the dust composition (Section~3.1). We then present the polarimetric properties of 3I/ATLAS (Section~3.2).


\subsection{Albedo and Thermal Properties of the Dust}\label{s:thermal-results}

\begin{figure}[htbp]
\centering
\includegraphics[width=\columnwidth]{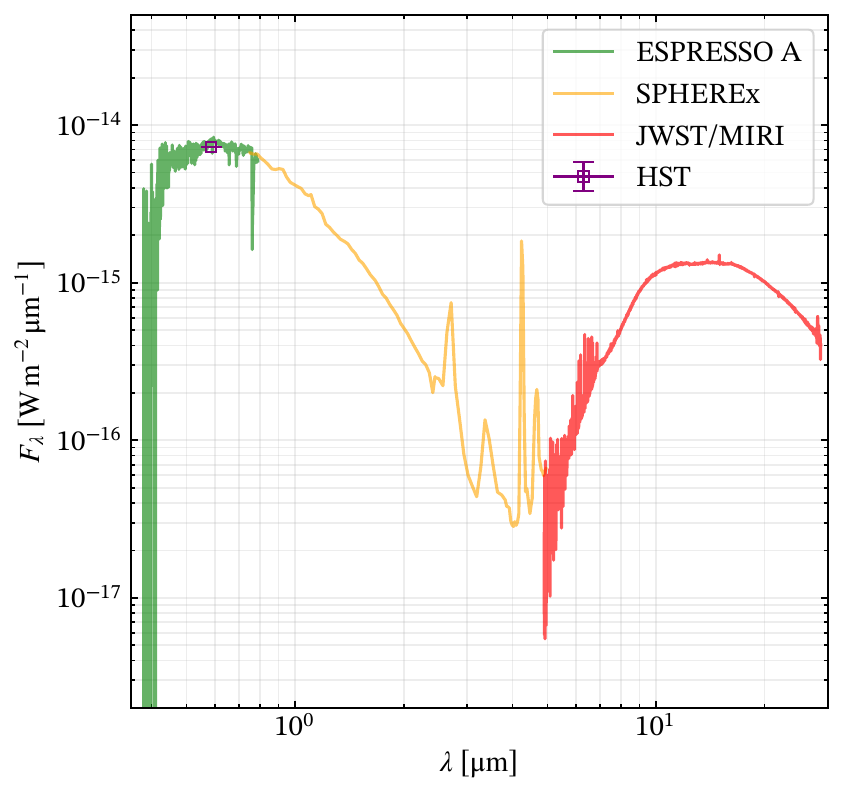}
\caption{Visible-to-mid-infrared SED of 3I/ATLAS, scaled to a $1\arcsec$-diameter photometric aperture at $r_{\rm H}=2.20$ au and $\Delta=1.80$ au. The HST photometry and VLT/ESPRESSO, SPHEREx, and JWST/MIRI MRS spectra are shown in purple, green, orange, and red, respectively.}
\label{fig:SED}
\end{figure}

Figure~\ref{fig:SED} shows the visible-to-mid-infrared SED of 3I/ATLAS. Despite spectral emission features in the near-infrared, a well-defined continuum extends from scattered sunlight to thermal emission. We model the SED in three steps: we first obtain a first-order estimate of the dust albedo without explicitly specifying the grain-size distribution or composition, then introduce a grain-size distribution and composition-dependent optical properties, and finally include a silicate component to reproduce the
$10~\mu\mathrm{m}$ spectral feature.


\subsubsection{Blackbody Estimate of the Dust Reflectance}
\label{s:albeo-BB}

\begin{figure}[htbp]
\centering
\includegraphics[width=\columnwidth]{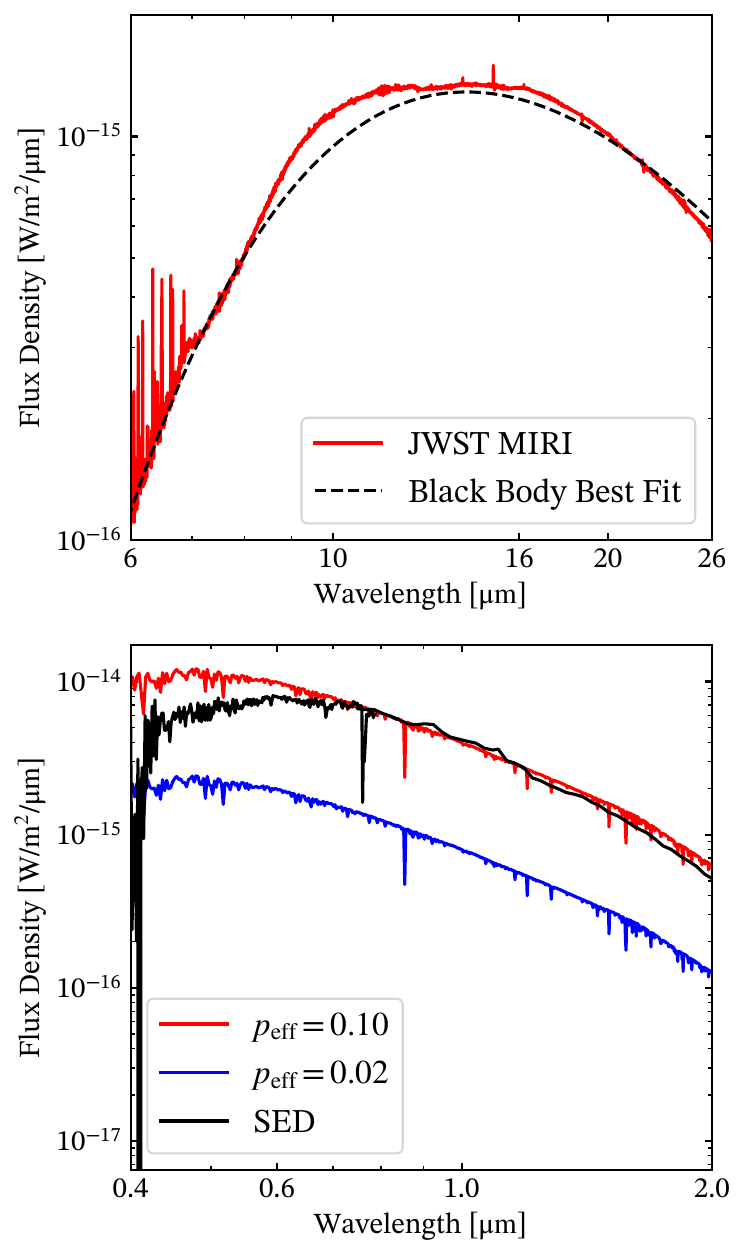}
\caption{Blackbody estimate of the dust reflectance. The upper panel shows the mid-infrared SED (red) and the fitted blackbody continuum (black dashed), excluding the spectral features reported by \citet{2026arXiv260627535B}. The lower panel compares the optical--near-infrared SED with scattered-light models for $p_{\rm eff}=0.02$ and 0.10, matched at wavelengths of $\lesssim0.4~\mu\mathrm{m}$ and $0.8$--$1.0~\mu\mathrm{m}$, respectively.
}
\label{fig:albedo-BB}
\end{figure}

As a first-order estimate without explicitly specifying the grain-size distribution or composition, we approximated the mid-infrared thermal emission by blackbody radiation. The formulation is given in Appendix~\ref{s:appendixC1}. In this approximation, the thermal emission provides an effective dust cross section, which is then used to estimate the reflectance from the scattered-light component. Fitting the JWST/MIRI continuum 
after masking the spectral features at 7.5–8.0 and 15–25 \micron\ yields a temperature of $206.9 \pm 0.01~\mathrm{K}$ and an effective solid angle of $8.63\times10^{-16} \pm 3.59\times10^{-17}~\mathrm{sr}$, corresponding to an equivalent emitting radius of $4.46\pm0.09~\mathrm{km}$. Using the thermally inferred cross section to scale the scattered-light component, we obtain a first-order estimate of the effective reflectance, $p_{\rm eff}\sim0.02$--$0.10$ (Figure~\ref{fig:albedo-BB}).  The inferred effective reflectance increases with wavelength, from approximately 0.02 at wavelengths shorter than $0.4~\micron$ to approximately 0.10 in the near-infrared ($0.8$--$2.0~\micron$).
The lower bound of \(p_{\rm eff}=0.02\) was set by the effective reflectance at the shortest wavelength with a reliable flux uncertainty.
Thus, independent of the detailed grain-size distribution and composition, the broad SED indicates a dark dust coma.

\subsubsection{Single-Component Grain-Size Distribution Models}
\label{s:single-model}

We next relaxed the blackbody approximation and modeled the dust as spherical, homogeneous, and isothermal grains following a power-law size distribution, $dN/da \propto a^{-q}$. We considered amorphous carbon, magnetite, graphite, astronomical silicate, amorphous olivine, and dirty water ice, for which suitable optical constants are available. For each grain size, we calculated the absorption efficiency using Mie theory and determined the equilibrium temperature from the energy balance between absorbed solar radiation and thermal emission, including sublimation cooling for water-ice-bearing grains following \citet{1986A&A...164..397M}. The scattered-light and thermal-emission components were then integrated over the grain-size distribution. Details of the dust model and fitting procedure are given in Appendices~\ref{s:appendixC2} and \ref{s:appendixC3}, respectively.

\begin{figure*}[htbp]
\centering
\includegraphics[width=\textwidth]{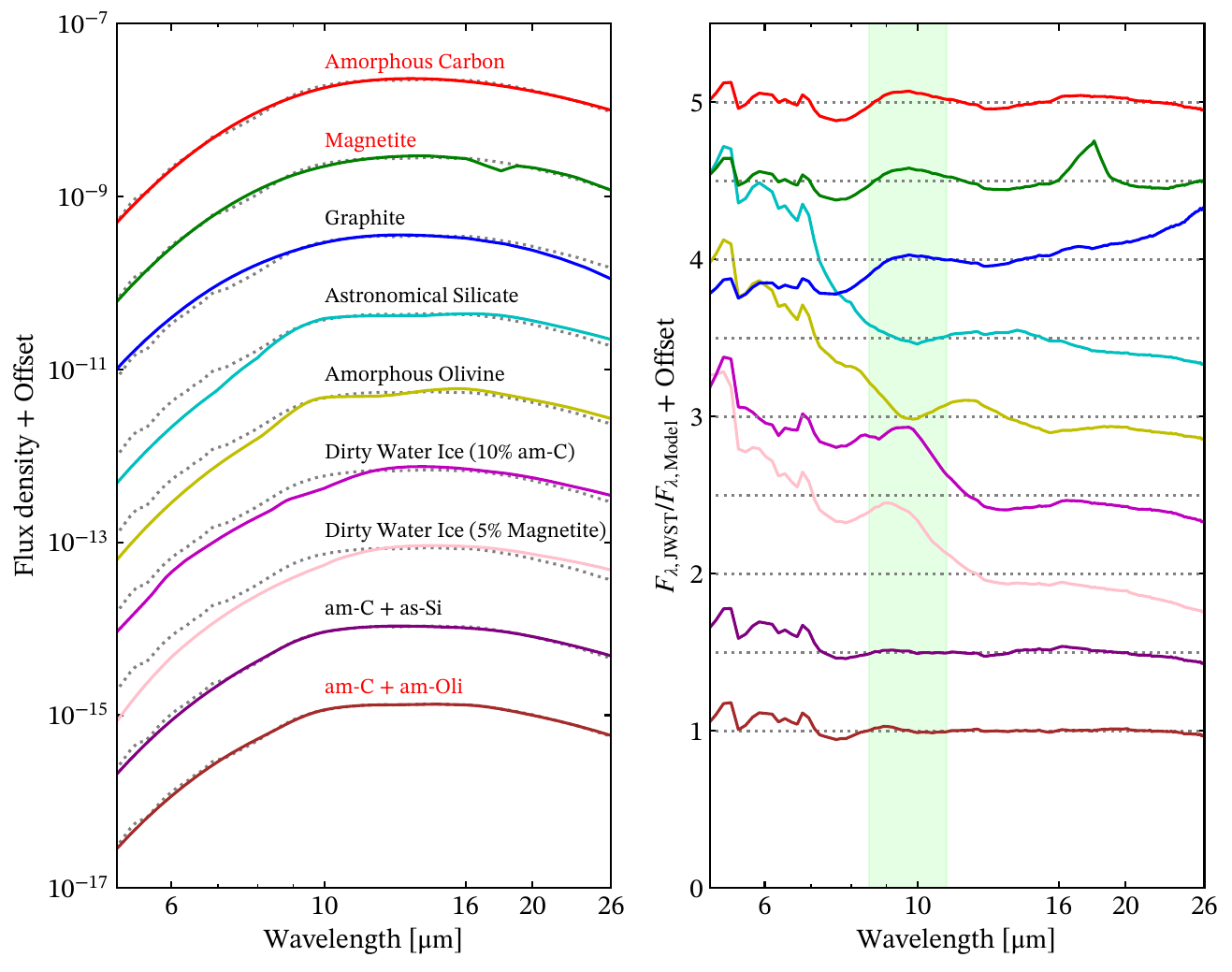}
\caption{
Thermal-emission models for the dust compositions considered here. The left panel compares the JWST/MIRI spectrum with the best-fitting models, and the right panel shows the corresponding observation-to-model flux ratios. Spectra and ratios are vertically offset for clarity. The shaded region marks the silicate emission feature reported by
\citet{2026arXiv260627535B}. In the two--component models, am--C and am--Oli denote amorphous carbon and amorphous olivine, respectively, while as--Si denotes astronomical silicate.
}
\label{fig:thermal_fit}
\end{figure*}

Figure~\ref{fig:thermal_fit} compares the JWST/MIRI spectrum with the best-fitting models for the compositions considered here. Strongly absorbing materials, particularly amorphous carbon and magnetite, reproduce the thermal continuum substantially better than weakly absorbing silicates and dirty water ice. For amorphous carbon and magnetite, the thermal emission near $6~\mathrm{\mu m}$ is reproduced by including small grains of $\sim1~\mathrm{\mu m}$ in the size distribution. Notably, the 15--25~$\mathrm{\mu m}$ continuum, which deviates from a simple blackbody \citep{2026arXiv260627535B}, is also reproduced by the grain-size distribution model without invoking an additional dust component. In contrast, the weakly absorbing amorphous olivine and astronomical silicate models, as well as the dirty-water-ice models, whose temperatures are reduced by sublimation cooling, cannot reproduce the observed thermal SED by adjusting the grain-size distribution alone. Magnetite, however, produces an emission feature near $17$--$18~\mathrm{\mu m}$ that is not apparent in the observed spectrum, while the graphite solution drives $q$ to its upper prior boundary. We therefore adopt amorphous carbon as the preferred single-component model.
For amorphous carbon, we obtain $q=3.24$, $a_{\min}=1.10~\mathrm{\mu m}$, and $a_{\max}\sim 23.2~\mathrm{\mu m}$, although the upper cutoff is only weakly constrained. 
The resulting effective V-band albedo is $0.041\pm0.001$, similar to that obtained for magnetite ($0.040\pm0.001$; Appendix~\ref{s:appendixC4}).
Thus, the composition-dependent models reinforce the first-order result that the SED is dominated by dark, strongly absorbing dust.

\subsubsection{Two-Component Dust Model}
\label{s:two-model}

While the single-component amorphous-carbon model reproduces the overall thermal continuum, residuals remain around the $10~\mathrm{\mu m}$ silicate feature reported by \citet{2026arXiv260627535B}. We therefore added a silicate component, considering astronomical silicate and amorphous olivine. The amorphous-carbon power-law index was fixed to the single-component value to reduce parameter degeneracy.

Among the models considered, the amorphous-carbon--amorphous-olivine model best reproduces the silicate feature (Figure~\ref{fig:thermal_fit}), with $q=3.04$ for amorphous olivine and minimum grain radii of order $1~\mathrm{\mu m}$. The upper size cutoff remains poorly constrained. The effective $V$-band albedo, $0.042\pm0.001$, is essentially unchanged from the single-component amorphous-carbon model. Thus, adding silicate improves the mid-infrared spectral structure without altering the inference of a low-albedo dust coma. Full fitting results are given in Appendix~\ref{s:appendixC3}.


\subsection{Polarimetric Properties}
\label{s:pol-results}

\begin{figure*}[htbp]
\centering
\includegraphics[width=\textwidth]{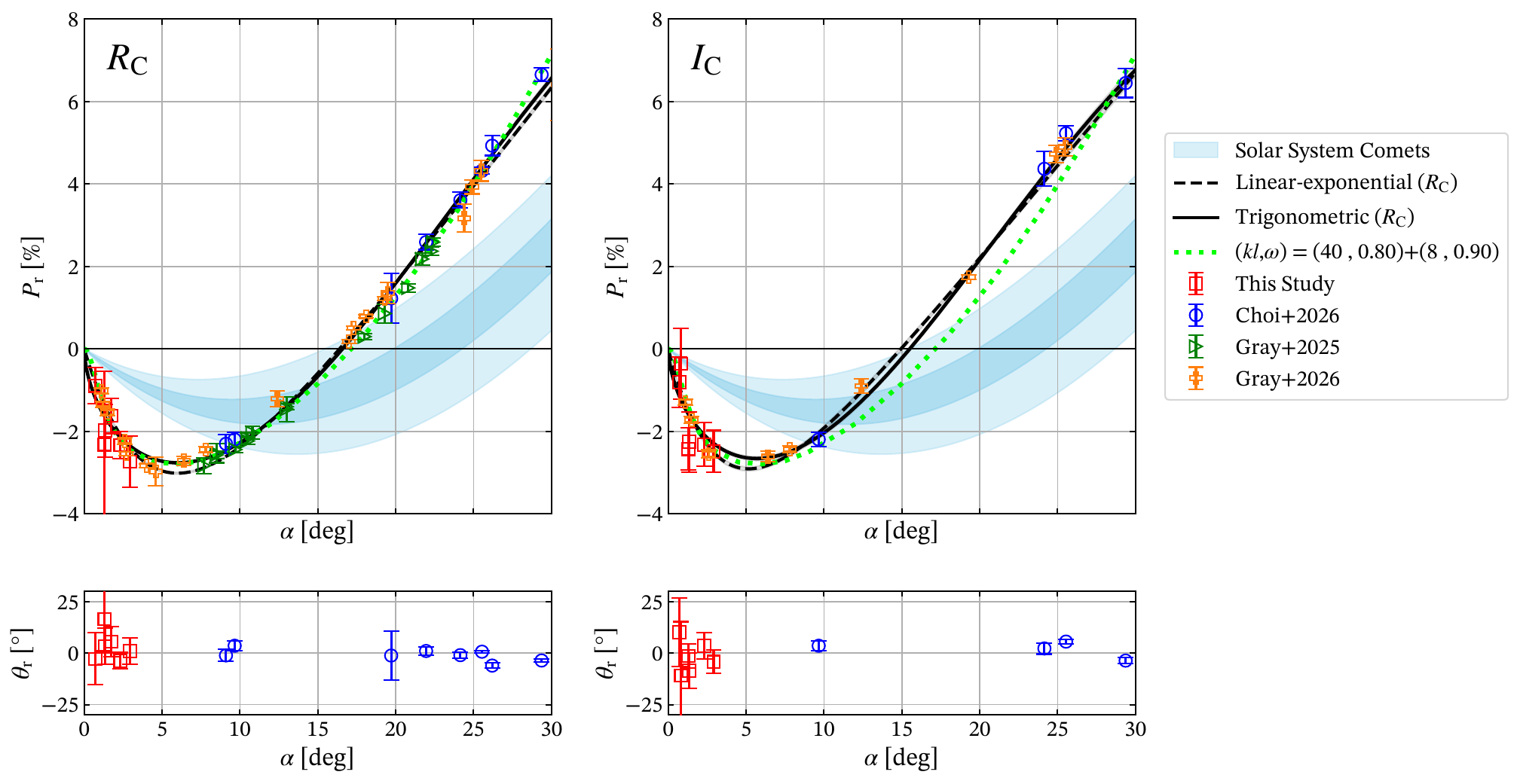}
\caption{
Polarization degree ($P_{\mathrm{r}}$) and polarization position angle ($\theta_{\mathrm{r}}$) of 3I/ATLAS as functions of phase angle, measured in the \Rc\ and \Ic\ bands. Because $\theta_{\mathrm{r}}$ approaches $90\degr$ in the negative-polarization branch, measurements with $\theta_{\mathrm{r}}>45\degr$ are displayed as $90\degr-\theta_{\mathrm{r}}$ for visual clarity. Red squares denote our MSI measurements, while green triangles, blue circles, and orange crosses denote previously published measurements by \citet{2025ApJ...992L..29G}, \citet{2026ApJ..1000L..29C}, and \citet{2026arXiv260904932G}, respectively. The sky-blue shaded regions indicate representative polarization phase curves of Solar System comets from the Database of Comet Polarimetry (DBCP; \citealt{Kiselev2017}). The black dashed and solid curves show the best-fitting linear--exponential and trigonometric empirical models, respectively, and the lime dotted curve indicates the best-fitting RT--CB model in the \Rc\ band.
}
\label{fig:PPC}
\end{figure*}


We fitted the combined post-perihelion \Rc-band polarization phase curve using the linear--exponential function \citep{Muinonen2002}, following the same parameterization adopted by \citet{2026arXiv260904932G}. The fit yields a polarization minimum of $P_{\min}=-3.04\pm0.028~\%$ at $\alpha_{\min}=5.98\pm0.054\degr$, an inversion angle of $\alpha_0=16.3\pm0.05\degr$, and a polarimetric slope of $h=0.439\pm0.004~\%/\degr$. These parameters are consistent with those reported by \citet{2026arXiv260904932G}, indicating that the overall low-phase-angle polarimetric behavior remained broadly unchanged across perihelion within the observational uncertainties. 

The negative-polarization minimum is substantially deeper than that typically observed for Solar System comets (Figure~\ref{fig:PPC}). Thus, although the polarization phase curve of 3I/ATLAS remained persistent across perihelion, its unusually deep negative-polarization branch represents a clear departure from the typical behavior of Solar System cometary dust. The physical implications of this behavior are discussed in Section~\ref{s:disc}.


\section{Discussion}\label{s:disc}


\subsection{Physical Constraints from the Polarization Phase Curve}


The unusually deep negative-polarization branch of 3I/ATLAS has motivated discussion of the physical properties and composition of its coma dust, including a possible contribution from water-ice grains. Based on the initial pre-perihelion measurements of \citet{2025ApJ...992L..29G}, \citet{2025MNRAS.544L.156Z} showed that the observed polarization could be reproduced by a dust model containing a substantial ($\sim30$--$40$ vol\%) pure-water-ice component. Possible changes in the contribution of icy grains around perihelion were subsequently discussed in connection with additional polarimetric measurements \citep{2025RNAAS...9..338Z,2026ApJ..1000L..29C}.

Our near-opposition measurements smoothly extend the pre- and post-perihelion polarization phase curves reported by \citet{2025ApJ...992L..29G,2026arXiv260904932G}, providing denser coverage around the polarization minimum and extending the measurements to $\alpha=0.70\degr$. A narrow polarimetric opposition effect has been observed at very small phase angles for icy bodies such as Europa \citep{2022PSJ.....3..134K}. Although our observations may not reach the minimum of such a feature, we find no indication that the negative polarization becomes deeper toward the smallest phase angles covered by our observations. The combined measurements confirm that the negative-polarization minimum is substantially deeper than that typically observed for Solar System comets. \citet{2026arXiv260904932G} obtained $P_{\min}=-2.94\pm0.03\%$ at $\alpha_{\min}=5.52\pm0.30^\circ$ in the \ensuremath{R}\ band, and attributed the difference from their pre-perihelion estimate primarily to their improved coverage at small phase angles rather than to dust evolution. Although our limited wavelength coverage does not allow us to quantify gas contamination as extensively as \citet{2026arXiv260904932G}, the close agreement of our \Rc- and \Ic-band measurements with their phase curves suggests that gas emission has little effect on our main polarimetric results.

Detailed modeling of cometary polarization with realistic dust aggregates requires exploration of a large parameter space involving particle size, optical constants, and aggregate structure \citep{2006A&A...449.1243K}. Rather than attempting such a detailed characterization, we use the publicly available RT–CB model of \citet{MARKKANEN2023108733} 
which is parameterized by
the single scattering albedo ($\tilde\omega$) and dimensionless extinction mean free path ($k\ell$), and as a simple first-order test of the scattering conditions required to reproduce our \Rc-band polarization phase curve. Using the polarization data alone, a two-component model provides a closer representation of the observed phase curve than the single-component models, with $(k\ell_1,\tilde{\omega}_1)=(40,0.80)$ and $(k\ell_2,\tilde{\omega}_2)=(8,0.90)$ and relative weights of approximately 0.61 and 0.39, respectively.

The RT–CB parameter $k\ell$ can be related to the volume filling factor $v$ of spherical scattering units through $(k\ell)\lambda/(2\pi)=4a/[3vQ_{\rm ext}(a,\lambda,m)]$ where $Q_{\rm ext}$ is the extinction efficiency, and $m$ is the
complex refractive index. \citet{2026ApJ..1000L..29C} inferred a characteristic submicron scattering scale for 3I/ATLAS; adopting $a\sim0.5~\mathrm{\mu m}$ and the optical properties of the absorbing materials favored by our SED analysis gives filling factors of only a few percent. Within the RT–CB framework, this low filling factor suggests a highly porous scattering structure. Interestingly, this first-order inference is qualitatively consistent with the more sophisticated aggregate modeling of \citet{2026arXiv260904932G}, who explored BCCA, BPCA, and more compact aggregate structures over a range of compositions and porosities and independently favored porous aggregates to reproduce the observed polarimetric properties.

\subsection{Dust Size Distribution, Composition, and Effective Albedo}


Our SED modeling provides independent constraints on the grain-size distribution of 3I/ATLAS. For the preferred single-component amorphous-carbon model, the differential grain-size distribution, $dN/da\propto a^{-q}$, has an index of $q=3.24$, comparable to the $q\simeq3.0$–$3.5$ range inferred for Solar System comets \citep{2007Icar..189..169I}. 
This value is also close to the $q = 3.5$ size distribution adopted by \citet{2026arXiv260904932G} in their modeling, although their light-scattering models are relatively insensitive to the exact value of $q$.
Our modeling indicates the presence of $\sim1~\micron$-sized grains, whereas dust-dynamical modeling and the absence of a prominent anti-solar tail suggest a minimum grain size of approximately $10~\micron$ for the dust dominating the large-scale coma morphology \citep{2026MNRAS.550g1164M}.
These results are not necessarily contradictory, because the infrared SED can be sensitive to a population of smaller grains that contributes little to the large-scale optical morphology.
In particular, the thermal continuum around $6~\micron$ requires a contribution from $\sim1~\micron$-sized grains, while the weak $10~\micron$ silicate feature further indicates the presence of small silicate grains in our model.
At the other end of the size distribution, the maximum grain size is poorly constrained. For sufficiently large grains, both the equilibrium temperature and the wavelength dependence of $Q_{\rm abs}$ become progressively less sensitive to grain size over the MIRI wavelength range, making $a_{\max}$ intrinsically difficult to determine from the present data. We therefore do not interpret the fitted $a_{\max}$ as a robust measurement of the physical maximum grain size.


The broad SED further indicates that the coma is dominated by dark, strongly absorbing dust. This conclusion does not uniquely require amorphous carbon; rather, amorphous carbon provides a representative description of the relatively featureless absorbing component that dominates the overall energy balance. In contrast, the dirty-water-ice models considered here do not reproduce the observed thermal emission, because the presence of ice substantially alters the grain temperature and thermal energy balance. This does not imply the absence of water ice in the coma: near-infrared observations have independently revealed absorption by solid H$_2$O \citep{2026ApJ..1000L..52L}. Instead, our SED modeling suggests that water ice does not dominate the thermally emitting dust population. The observed ice absorption may arise from a minor ice-bearing component or from ice grains physically distinct from the strongly absorbing dust dominating the thermal continuum. This picture is also compatible with the polarimetric modeling of \citet{2026arXiv260904932G}, in which water ice alone could not simultaneously reproduce the observed polarization phase curve and optical color. Likewise, the addition of amorphous silicate improves the representation of the $10~\mathrm{\mu m}$ emission feature identified by \citet{2026arXiv260627535B}, but does not substantially alter the overall energy balance or the inferred optical albedo.


Both the simple blackbody estimate and the single- and two-component grain-size distribution models indicate a low effective optical albedo. The preferred models give a $V$-band value of approximately $0.04$–$0.05$, broadly consistent with the low albedos of Solar System cometary dust \citep{2003JQSRT..79..695H,2010ApJ...714.1324I}.
A substantially higher geometric albedo was obtained for the aggregate particles modeled by \citet{2026arXiv260904932G} from polarimetric and optical-colour constraints. This difference may partly reflect the different sensitivities of the two approaches to multiple dust components. The high-albedo particles inferred from polarimetric modeling may therefore represent a dust component that contributes particularly strongly to the observed polarization, while other dust components may also contribute to the overall coma. A mixture of bright, weakly absorbing particles and dark, strongly absorbing particles could reconcile the two results. Exploring such a multi-component picture may provide an interesting direction for future studies.
Combining the scattered-light and thermal-emission components of the SED provides a direct observational constraint on the effective dust albedo, following an established approach for cometary dust. Because the dust albedo of 3I/ATLAS has largely been assumed rather than directly measured in previous studies, our estimate provides an important reference for future determinations of dust cross sections and mass-loss rates from optical observations.

Taken together, these results reveal an interesting contrast. The effective albedo and grain-size distribution of 3I/ATLAS inferred from our SED analysis are broadly comparable to those of Solar System comets, whereas its polarization phase curve is distinctly different, exhibiting an exceptionally deep negative-polarization branch. \citet{2026ApJ..1000L..29C} also suggested the presence of submicron constituent particles from the wavelength dependence of the polarization, which shows a possible peak in the near-infrared. Together with this constraint, both our simple RT–CB analysis and the more detailed aggregate modeling of \citet{2026arXiv260904932G} point toward a porous scattering structure. These results suggest that aggregate structure may play an important role in the unusual polarization of 3I/ATLAS.

Viewed from another perspective, 3I/ATLAS also highlights the remarkable similarity of the polarization phase curves observed among Solar System comets. Because cometary polarization depends on multiple dust properties, including composition, particle size, and aggregate structure, the relatively limited diversity of their polarization phase curves may reflect common properties inherited from the proto-solar nebula and/or limited alteration of refractory dust after formation. In this sense, an interstellar visitor such as 3I/ATLAS provides not only a probe of material formed around another star, but also a new perspective on the apparent uniformity of cometary dust within our own Solar System.


\begin{acknowledgments}
This research was supported by a grant from the Korean National Research Foundation (NRF) (MEST), funded by the Korean government (No. 2023R1A2C1006180).
\end{acknowledgments}

\bibliography{sample701}{}
\bibliographystyle{aasjournalv7}

\appendix
\twocolumngrid
\section{Data Reduction on Imaging Polarimetry}\label{s:appendixA}
The preprocessing included bias subtraction, dark subtraction, flat-field correction, and cosmic-ray rejection. After preprocessing, aperture photometry was performed using a circular aperture with a radius of 5~arcsec, chosen to include as much of the coma signal as possible while maintaining adequate signal-to-noise ratio. At the geocentric distance of 3I/ATLAS at the time of observation, this aperture corresponds to a projected photocentric radius of approximately 9000~km. The local sky background was estimated using an annulus with inner and outer radii of 7.5 and 11~arcsec, respectively, corresponding to projected photocentric distances of approximately 13,000 and 19,000~km. The normalized Stokes parameters were derived following the same method as \citet{2022MNRAS.509.4128I}. The measured normalized Stokes parameters were corrected for instrumental polarization ($q_{\rm inst}$ and $u_{\rm inst}$), polarimetric efficiency ($\eta_\mathrm{eff}$), and the polarization position-angle offset ($\theta_{\rm offset}$) using the calibration coefficients listed in Table~\ref{tab:polcoeff}. For the \Ic\ band, we assumed the same effective polarization efficiency, \(\eta_\mathrm{eff}\), as in the \Rc\ band because no suitable polarization calibrator was available at \Ic-band wavelengths. The resulting aperture-polarimetry measurements in the \Rc\ and \Ic\ bands are summarized in Table~\ref{tab:sumpol}.

\setcounter{table}{0}
\renewcommand{\thetable}{A\arabic{table}}
\begin{deluxetable*}{ccccc}[h]
\label{tab:polcoeff}
\tabletypesize{\small}
\tablewidth{\textwidth}
\tablecolumns{5}
\setlength{\tabcolsep}{6pt}
\tablecaption{Instrumental polarization calibration coefficients of MSI.}
\tablehead{
\colhead{Filter} & \colhead{$q_\text{inst}$ [\%]} &
\colhead{$u_\text{inst}$ [\%]} & \colhead{$\eta_\text{eff}$ [\%]} &
\colhead{$\theta_\text{offset}$ [$\degr$]}
}
\startdata
\Rc & 0.645 $\pm$ 0.275 & 0.407 $\pm$ 0.094 & 99.09 $\pm$ 0.02 & 2.32 $\pm$ 0.43 \\
\Ic & 0.098 $\pm$ 0.140 & -0.126 $\pm$ 0.161 & 99.09 $\pm$ 0.02 & 1.43 $\pm$ 0.32 \\\hline
\enddata

\end{deluxetable*}

\begin{deluxetable*}{@{\extracolsep{\fill}}cccccccccccc@{}}[t]
\tabletypesize{\small}
\tablewidth{\textwidth}
\tablecolumns{12}

\tablecaption{Optical polarimetric measurements of 3I/ATLAS.
\label{tab:sumpol}}

\tablehead{
\colhead{UT Date} &
\colhead{Filter} &
\colhead{$r_{\rm H}$} &
\colhead{$\alpha$} &
\colhead{$P$} &
\colhead{$\sigma_P$} &
\colhead{$\theta$} &
\colhead{$\sigma_\theta$} &
\colhead{$P_{\rm r}$} &
\colhead{$\sigma_{P_{\rm r}}$} &
\colhead{$\theta_{\rm r}$} &
\colhead{$\sigma_{\theta_{\rm r}}$} \\
\colhead{} &
\colhead{} &
\colhead{[au]} &
\colhead{[\degr]} &
\colhead{[\%]} &
\colhead{[\%]} &
\colhead{[\degr]} &
\colhead{[\degr]} &
\colhead{[\%]} &
\colhead{[\%]} &
\colhead{[\degr]} &
\colhead{[\degr]} \\
\colhead{(1)} &
\colhead{(2)} &
\colhead{(3)} &
\colhead{(4)} &
\colhead{(5)} &
\colhead{(6)} &
\colhead{(7)} &
\colhead{(8)} &
\colhead{(9)} &
\colhead{(10)} &
\colhead{(11)} &
\colhead{(12)}
}

\startdata
2026-01-17.63 & \Rc & 3.192 & 2.945 & 2.803 & 0.621 & -63.18 & 6.35 & -2.733 & 0.621 & 89.13 & 6.35 \\
2026-01-17.60 & \Ic & 3.191 & 2.913 & 2.562 & 0.509 & -57.92 & 5.69 & -2.484 & 0.509 & 94.25 & 5.69 \\
2026-01-18.72 & \Rc & 3.226 & 2.325 & 2.374 & 0.324 & -55.14 & 3.91 & -2.333 & 0.324 & 93.67 & 3.91 \\
2026-01-18.69 & \Ic & 3.225 & 2.303 & 2.397 & 0.530 & -62.13 & 6.33 & -2.321 & 0.530 & 86.53 & 6.33 \\
2026-01-20.47 & \Rc & 3.281 & 1.351 & 2.092 & 0.637 & -48.96 & 8.73 & -1.980 & 0.637 & 86.70 & 8.73 \\
2026-01-20.50 & \Ic & 3.283 & 1.339 & 2.465 & 0.729 & -36.85 & 8.48 & -2.252 & 0.729 & 98.50 & 8.48 \\
2026-01-22.42 & \Rc & 3.345 & 0.701 & 0.997 & 0.438 & 9.57 & 12.60 & -0.891 & 0.438 & 92.83 & 12.60 \\
2026-01-22.43 & \Ic & 3.345 & 0.700 & 1.052 & 0.610 & -3.06 & 16.62 & -0.805 & 0.610 & 79.89 & 16.62 \\
2026-01-23.40 & \Ic & 3.377 & 0.826 & 0.938 & 0.856 & 56.25 & 26.15 & -0.357 & 0.856 & 100.94 & 26.15 \\
2026-01-24.70 & \Rc & 3.418 & 1.309 & 3.286 & 1.780 & 53.77 & 15.53 & -2.316 & 1.780 & 73.50 & 15.53 \\
2026-01-24.65 & \Ic & 3.417 & 1.280 & 2.481 & 0.507 & 70.86 & 5.85 & -2.426 & 0.507 & 91.39 & 5.85 \\
2026-01-25.65 & \Rc & 3.449 & 1.720 & 1.705 & 0.428 & 72.82 & 7.20 & -1.620 & 0.428 & 84.50 & 7.20 \\
\enddata
\tablenotetext{}{NOTE--(1) UT Date. (2) Filter used in each observation. (3) Heliocentric distance. (4) Phase angle. (5) Observed polarization degree. (6) Standard deviation of $P$. (7) Position angle of the strongest electric field vector. (8) Standard deviation of $\theta$. (9) Polarization degree relative to the scattering plane. (10) Standard deviation of $P_\mathrm{r}$ (11) Position angle of the strongest electric field vector relative to the scattering plane normal. (12) Standard deviation of $\theta_\mathrm{r}.$}
\end{deluxetable*}

\section{SED generation}\label{s:appendixB}

\setcounter{table}{0}
\renewcommand{\thetable}{B\arabic{table}}

\begin{deluxetable}{lcccc}
\tabletypesize{\small}
\tablewidth{\columnwidth}
\tablecaption{Summary of the observations used in this study.
\label{tab:B1}}
\tablehead{
\colhead{Facility/Instrument} &
\colhead{UT Date} &
\colhead{$r_{\mathrm{H}}$} &
\colhead{$\Delta_\mathrm{obs}$} &
\colhead{$\alpha$}\\
\colhead{} &
\colhead{} &
\colhead{[au]} &
\colhead{[au]} &
\colhead{[\degr]}
}
\startdata
HST           & 2025 Dec 12     & 2.11 & 1.81 & 27.8 \\
JWST/MIRI     & 2025 Dec 15     & 2.18 & 1.80 & 26.8 \\
SPHEREx       & 2025 Dec 8--15  & 2.07 & 1.82 & 28.4 \\
VLT/ESPRESSO  & 2025 Dec 21     & 2.35 & 1.80 & 23.0 \\
\enddata
\tablecomments{
$r_{\mathrm{H}}$, $\Delta_{\mathrm{obs}}$, and $\alpha$ denote the heliocentric distance, observer--target distance, and solar phase angle of 3I/ATLAS, respectively. For the SPHEREx observations, the listed values were evaluated at the midpoint of the observing interval (2025 December 11.5 UT).
}
\end{deluxetable}

To construct the optical-to-mid-infrared SED of 3I/ATLAS, we combined archival observations obtained with JWST/MIRI MRS, SPHEREx, VLT/ESPRESSO, and HST/WFC3. The instrumental configurations, observing strategies, and standard data-reduction procedures for these datasets have been described in the corresponding publications \citep{2026ApJ..1001L..11B,2026RNAAS..10...26L, 2026arXiv260715355L}. Here, we describe only the additional processing and cross-instrument scaling applied in this work.

\subsection{JWST/MIRI MRS}

We retrieved the Stage~3 MIRI/MRS spectral cubes of 3I/ATLAS and the corresponding off-target background observations obtained on 2025 December 15 under JWST DDT Program 9442 (PI: M. Belyakov). We adopted the data-selection criteria of \citet{2026ApJ..1001L..11B} and performed no additional detector-level reprocessing.

Source and background spectra were extracted from the \texttt{*s3d.fits} cubes using circular apertures with a radius of $0.5\arcsec$, corresponding to a projected photocentric radius of approximately 650~km. After background subtraction, the individual MRS spectral segments were combined into a continuous spectrum. Small discontinuities between overlapping spectral segments were corrected using multiplicative scale factors derived from the overlap regions, with Channel~3 SHORT adopted as the flux reference.

\subsection{HST/WFC3}

We used archival HST/WFC3 imaging obtained on 2025 December 12 with the F350LP filter. Aperture photometry was performed using an aperture matched to the projected physical radius adopted for the JWST/MIRI extraction. The angular aperture size was therefore adjusted for the different observer--object distance at the HST epoch.

The local background was estimated from a distant annulus where the radial surface-brightness profile of the coma became indistinguishable from the surrounding sky. The resulting F350LP photometry was adopted as the absolute optical-flux reference for the construction of the combined SED. In particular, it was used to determine the multiplicative correction applied to the ESPRESSO continuum.

\subsection{VLT/ESPRESSO}

For the optical spectrum, we used the publicly available, nightly-stacked fiber-A spectrum obtained on 2025 December 21 (epoch E23-1) \citep{2026arXiv260715355L,luco_2026_20966879}. The spectrum had already been reduced and flux calibrated with the standard ESPRESSO pipeline.

Because the fiber sampled the extended coma and an independent local sky measurement was not available for all exposures, residual systematic uncertainty remained in the absolute continuum level. We therefore applied an additional multiplicative normalization by matching the ESPRESSO continuum to the HST/WFC3 F350LP photometry. Thus, the HST measurement provides the absolute optical-flux anchor for the ESPRESSO spectrum used in our SED analysis.

Prior to the comparison and subsequent SED modeling, the ESPRESSO spectrum was rebinned using an inverse-variance-weighted mean.

\subsection{SPHEREx}

We adopted the SPHEREx spectrophotometry of 3I/ATLAS obtained between 2025 December 8 and 15 from \citet{2026RNAAS..10...26L}. Because the SPHEREx photometric aperture is substantially larger than the $0.5\arcsec$-radius aperture adopted for JWST/MIRI, an additional multiplicative correction was required before combining the datasets.

We determined this correction from the continuum region shared by SPHEREx and the HST-normalized ESPRESSO spectrum at $0.75$--$0.79~\micron$. The median SPHEREx-to-ESPRESSO continuum ratio yields a multiplicative scale factor of 0.070. Consequently, the SPHEREx normalization is indirectly tied to the HST photometric reference through the corrected ESPRESSO spectrum.

For the dust-continuum analysis, we retained only SPHEREx channels that are not significantly affected by identified gas emission, terrestrial exospheric emission, or possible residual broad ice-absorption features. The wavelength intervals used in the SED modeling are $0.75$--$0.80~\micron$ and $1.2$--$2.4~\micron$.

Because the observations were obtained at different epochs, all datasets were corrected for differences in observing geometry before constructing the final SED. The corresponding observing circumstances and scaling factors are summarized in Table~\ref{tab:B1}.

\section{Dust Model}\label{s:appendixC}

\setcounter{table}{0}
\renewcommand{\thetable}{C\arabic{table}}

\setcounter{figure}{0}
\renewcommand{\thefigure}{C\arabic{figure}}

This Appendix summarizes the equations, assumptions, and fitting parameters used for the blackbody and composition-dependent dust models.

\subsection{Blackbody Reflectance Estimate}
\label{s:appendixC1}

As a first-order estimate of the optical reflectance, the mid-infrared thermal emission was approximated by a blackbody,
\begin{equation}
F_{\lambda,\rm th} = \Omega_{\rm eff} B_\lambda(T),
\label{eq:C_BB}
\end{equation}
where $F_{\lambda,\rm th}$ is the thermal spectral flux density, $B_\lambda(T)$ is the Planck function at temperature $T$, and $\Omega_{\rm eff}$ is the effective solid angle. The parameters $T$ and $\Omega_{\rm eff}$ were determined by Markov Chain Monte Carlo (MCMC) fitting \citep{2013PASP..125..306F}. The effective solid angle is related to the emitting cross section by
\begin{equation}
\Omega_{\rm eff} = \frac{C_{\rm dust}}{\Delta_{\rm obs}^2} = \frac{\pi a_{\rm eff}^2}{\Delta_{\rm obs}^2},
\label{eq:C_omega}
\end{equation}
where $C_{\rm dust}$ is the effective dust cross section, $\Delta_{\rm obs}$ is the observer--target distance, and$a_{\rm eff}$ is an equivalent cross-sectional radius; it does not represent the physical radius of the coma or of an individual grain. Assuming that the same effective cross section contributes to the scattered-light component,
\begin{equation}
F_{\lambda,\rm sca}
= \frac{S_{\odot,\lambda}}{r_{\rm H}^2}
\frac{\pi a_{\rm eff}^2}{\Delta_{\rm obs}^2}
p_{\rm eff}(\lambda),
\label{eq:C_albedo}
\end{equation}
where $F_{\lambda,\rm sca}$ is the scattered-light spectral flux density, $S_{\odot,\lambda}$ is the solar spectral irradiance at 1 au, $r_{\rm H}$ is the heliocentric distance in au, and $p_{\rm eff}(\lambda)$ is the wavelength-dependent effective albedo.

The blackbody fit was performed over $6.0$--$26~\micron$, excluding $7.5$--$8.0~\micron$ and $15$--$25~\micron$ affected by prominent spectral features.


\subsection{Single--Composition Grain--Size Distribution Model}
\label{s:appendixC2}
We modeled the dust coma as optically thin and composed of spherical, homogeneous, and isothermal grains. The observed SED was calculated as the sum of scattered sunlight and thermal emission,
\begin{equation}
\label{eq:C_SED}
\begin{aligned}
F_\lambda = \frac{1}{\Delta_{\rm obs}^2} \Bigg[ & \frac{S_\odot(\lambda)}{r_{\rm H}^2} p(\lambda) \int_{a_{\min}}^{a_{\max}} a^2 \frac{dN}{da}\,da
\\
&+ \int_{a_{\min}}^{a_{\max}} Q_{\rm abs}(a,\lambda) B_\lambda[T_{\rm eq}(a)] \pi a^2 \frac{dN}{da}\,da \Bigg],
\end{aligned}
\end{equation}
where $F_\lambda$ is the total spectral flux density, $p(\lambda)$ is the wavelength-dependent effective reflectance, $Q_{\rm abs}(a,\lambda)$ is the absorption efficiency of a grain of radius $a$, $T_{\rm eq}(a)$ is the equilibrium temperature of the grain, and $a_{\min}$ and $a_{\max}$ are the minimum and maximum grain radii, respectively. $N$ means the number of dust grains.
The grain-size distribution was assumed to follow
\begin{equation}
\label{eq:C_size}
\frac{dN}{da} = N_0\ \left(\frac{a}{1~\micron}\right)^{-q},
\qquad a_{\min}\leq a\leq a_{\max},
\end{equation}
where $N_0$ is the normalization factor and $q$ is the power-law index.

The equilibrium temperature was obtained from the energy-balance condition
\begin{equation}
E_{\rm abs}=E_{\rm th}+E_{\rm sub},
\label{eq:C_energy}
\end{equation}
with
\begin{equation}
E_{\rm abs} = \pi a^2 \left(\frac{1~{\rm au}}{r_{\rm H}}\right)^2 \int S_\odot(\lambda) Q_{\rm abs}(a,\lambda) \,d\lambda,
\end{equation}
and
\begin{equation}
E_{\rm th} = 4\pi a^2 \int \pi B_\lambda(T_{\rm eq}) Q_{\rm abs}(a,\lambda) \,d\lambda,
\end{equation}

\noindent where $E_{\rm abs}$, $E_{\rm th}$, and $E_{\rm sub}$ are the absorbed
solar power, thermally emitted power, and energy loss due to sublimation,
respectively.
For refractory grains, $E_{\rm sub}=0$. For water-ice-bearing grains, the sublimation-cooling term was calculated following \citet{1986A&A...164..397M}, adopting the molecular weight $\mu_{\rm ice}=18$, the bulk density $\rho_{\rm ice}=1.0~{\rm g\,cm^{-3}}$ , and the temperature-dependent latent heat of water ice from \citet{DELSEMME19711229}.

The absorption efficiencies were calculated using Mie theory with \texttt{miepython}. We adopted the optical constants of magnetite, graphite, astronomical silicate, and water ice from \citet{1989eidr.proc..397M}, amorphous carbon from \citet{1993A&A...279..577P}, and amorphous olivine from \citet{1995A&A...300..503D}. For the dirty-water-ice models, we used water ice containing either 5\% magnetite or 10\% amorphous carbon by volume; the former was calculated using the Maxwell--Garnett mixing rule. The solar input spectrum was the TSIS-1 Hybrid Solar Reference Spectrum \citep{2021GeoRL..4891709C}.


\subsection{Thermal Fitting and Optical Albedo}
\label{s:appendixC3}
For each single-component model, the free parameters were $N_0$, $q$, $a_{\min}$, and $a_{\max}$. Uniform priors were adopted over $1<q<6$, $ 0.01<a_{\min}<10~\micron$, and $10<a_{\max}<10^6~\micron$.

For the two-component models, the total spectrum was calculated as the sum of the two dust populations. We considered amorphous carbon combined with either astronomical silicate or amorphous olivine. The amorphous-carbon power-law index was fixed at the single-component best-fitting value, $q=3.241$, while the remaining parameters were fitted using the same priors.

The goodness of fit was evaluated using
\begin{equation}
\label{eq:C_chi}
\chi^2_{\rm red} = \frac{1}{\nu} \sum_{i\in{\rm fit}} \left[ \frac{ F_{\lambda,i}^{\rm model} - F_{\lambda,i}^{\rm JWST}}{\sigma_{\lambda,i}^{\rm JWST}}\right]^2,
\end{equation}
where $\nu$ is the number of degrees of freedom, $F_{\lambda,i}^{\rm model}$ and $F_{\lambda,i}^{\rm JWST}$ are the model and observed JWST flux densities in the $i$-th spectral bin, respectively, and $\sigma_{\lambda,i}^{\rm JWST}$ is the corresponding uncertainty in the JWST flux density. The numbers of degrees of freedom were 164 and 161 for the single- and two-component models, respectively. The posterior constraints are summarized in Table~\ref{tab:C1}.

After constraining the thermal model, the wavelength-dependent effective albedo was derived from the scattered-light term in Equation~(\ref{eq:C_SED}). The representative $V$-band albedo was estimated by averaging $p(\lambda)$ over $0.4$--$0.6~\micron$.

\begin{sidewaystable*}
\centering
\caption{Thermal-emission MCMC parameters. Values are posterior medians; errors are 16th--84th percentile offsets. A dagger marks a MAP parameter at a prior boundary.}
\label{tab:C1}
\begin{tabular}{lccccccccc}
\hline
Material & \multicolumn{2}{c}{$N_0$ [$\mu$m$^{-1}$]} & \multicolumn{2}{c}{$q$} & \multicolumn{2}{c}{$a_{\min}$ [$\mu$m]} & \multicolumn{2}{c}{$a_{\max}$ [$\mu$m]} & $\chi^2_\nu$ \\
 & Value & Error & Value & Error & Value & Error & Value & Error & \\
\hline
Amorphous carbon & $1.47\times10^{19}$ & $\begin{array}{r}+8.46\times10^{15}\\-8.44\times10^{15}\end{array}$ & $3.241$ & $\begin{array}{r}+2.05\times10^{-3}\\-1.93\times10^{-3}\end{array}$ & $1.104$ & $\begin{array}{r}+1.11\times10^{-4}\\-1.11\times10^{-4}\end{array}$ & $23.17$ & $\begin{array}{r}+0.3001\\-0.2746\end{array}$ & $8113$ \\
Magnetite & $9.36\times10^{18}$ & $\begin{array}{r}+5.36\times10^{15}\\-5.47\times10^{15}\end{array}$ & $3.024$ & $\begin{array}{r}+8.96\times10^{-4}\\-8.5\times10^{-4}\end{array}$ & $0.9441$ & $\begin{array}{r}+1.35\times10^{-4}\\-1.38\times10^{-4}\end{array}$ & $33.33$ & $\begin{array}{r}+0.2432\\-0.2298\end{array}$ & $21490$ \\
Graphite & $2.08\times10^{21}$ & $\begin{array}{r}+9.54\times10^{17}\\-9.71\times10^{17}\end{array}$ & $6$\textsuperscript{\ensuremath{\dagger}} & $\begin{array}{r}+4.66\times10^{-6}\\-1.03\times10^{-5}\end{array}$ & $2.828$ & $\begin{array}{r}+3.92\times10^{-4}\\-3.97\times10^{-4}\end{array}$ & $10$\textsuperscript{\ensuremath{\dagger}} & $\begin{array}{r}+9.2\times10^{-5}\\-3.96\times10^{-5}\end{array}$ & $27763$ \\
Astronomical silicate & $2.77\times10^{18}$ & $\begin{array}{r}+5.81\times10^{14}\\-5.91\times10^{14}\end{array}$ & $3.027$ & $\begin{array}{r}+4.05\times10^{-5}\\-4.14\times10^{-5}\end{array}$ & $0.01001$\textsuperscript{\ensuremath{\dagger}} & $\begin{array}{r}+1.74\times10^{-5}\\-7.52\times10^{-6}\end{array}$ & $1\times10^{6}$\textsuperscript{\ensuremath{\dagger}} & $\begin{array}{r}+6.74\\-15.23\end{array}$ & $28215$ \\
Amorphous olivine & $1\times10^{18}$ & $\begin{array}{r}+2.76\times10^{14}\\-2.78\times10^{14}\end{array}$ & $2.886$ & $\begin{array}{r}+3.55\times10^{-5}\\-3.59\times10^{-5}\end{array}$ & $0.01$\textsuperscript{\ensuremath{\dagger}} & $\begin{array}{r}+2.65\times10^{-6}\\-1.24\times10^{-6}\end{array}$ & $1\times10^{6}$\textsuperscript{\ensuremath{\dagger}} & $\begin{array}{r}+3.895\\-8.255\end{array}$ & $49497$ \\
DWI (aC) & $1.97\times10^{18}$ & $\begin{array}{r}+9.67\times10^{14}\\-9.26\times10^{14}\end{array}$ & $2.965$ & $\begin{array}{r}+7.22\times10^{-5}\\-6.91\times10^{-5}\end{array}$ & $0.05381$ & $\begin{array}{r}+7.37\times10^{-4}\\-8.48\times10^{-4}\end{array}$ & $1\times10^{6}$\textsuperscript{\ensuremath{\dagger}} & $\begin{array}{r}+5.685\\-12.57\end{array}$ & $118251$ \\
DWI (magnetite) & $9.24\times10^{17}$ & $\begin{array}{r}+5.17\times10^{14}\\-5.59\times10^{14}\end{array}$ & $2.867$ & $\begin{array}{r}+6.6\times10^{-5}\\-7.14\times10^{-5}\end{array}$ & $0.01004$\textsuperscript{\ensuremath{\dagger}} & $\begin{array}{r}+6.27\times10^{-5}\\-2.79\times10^{-5}\end{array}$ & $1\times10^{6}$\textsuperscript{\ensuremath{\dagger}} & $\begin{array}{r}+5.728\\-13.51\end{array}$ & $122494$ \\
\hline
\multicolumn{9}{l}{Amorphous carbon + astronomical silicate} & $2654$ \\
\quad Amorphous carbon & $1.18\times10^{19}$ & $\begin{array}{r}+9.11\times10^{15}\\-9.34\times10^{15}\end{array}$ & $3.241$ (fixed) & -- & $1.1$ & $\begin{array}{r}+1.55\times10^{-4}\\-1.52\times10^{-4}\end{array}$ & $72.52$ & $\begin{array}{r}+0.4899\\-0.4724\end{array}$ &  \\
\quad Astronomical silicate & $6.93\times10^{14}$ & $\begin{array}{r}+1.58\times10^{12}\\-9.8\times10^{11}\end{array}$ & $6$\textsuperscript{\ensuremath{\dagger}} & $\begin{array}{r}+2.43\times10^{-4}\\-5.17\times10^{-4}\end{array}$ & $0.01$\textsuperscript{\ensuremath{\dagger}} & $\begin{array}{r}+4\times10^{-6}\\-1.88\times10^{-6}\end{array}$ & $3.36\times10^{3}$ & $\begin{array}{r}+1.62\times10^{5}\\-3.3\times10^{3}\end{array}$ &  \\
\multicolumn{9}{l}{Amorphous carbon + amorphous olivine} & $806$ \\
\quad Amorphous carbon & $1.03\times10^{19}$ & $\begin{array}{r}+7.01\times10^{15}\\-7.1\times10^{15}\end{array}$ & $3.241$ (fixed) & -- & $1.066$ & $\begin{array}{r}+1.67\times10^{-4}\\-1.68\times10^{-4}\end{array}$ & $10$\textsuperscript{\ensuremath{\dagger}} & $\begin{array}{r}+4.08\times10^{-4}\\-1.8\times10^{-4}\end{array}$ &  \\
\quad Amorphous olivine & $1.14\times10^{18}$ & $\begin{array}{r}+2.4\times10^{15}\\-2.39\times10^{15}\end{array}$ & $3.037$ & $\begin{array}{r}+3.77\times10^{-4}\\-3.81\times10^{-4}\end{array}$ & $0.6713$ & $\begin{array}{r}+1.44\times10^{-3}\\-1.43\times10^{-3}\end{array}$ & $9.98\times10^{5}$\textsuperscript{\ensuremath{\dagger}} & $\begin{array}{r}+1.73\times10^{3}\\-3.89\times10^{3}\end{array}$ &  \\
\hline
\end{tabular}
\begin{flushleft}\footnotesize $N_0$ is defined by $dN/da=N_0(a/1\,\mu\mathrm{m})^{-q}$. Priors are $1<q<6$, $0.01<a_{\min}<10\,\mu$m, and $10<a_{\max}<10^6\,\mu$m.\end{flushleft}
\end{sidewaystable*}

\subsection{Effective Reflectance Result}
\label{s:appendixC4}
\begin{figure}[htbp]
\centering
\includegraphics[width=\columnwidth]{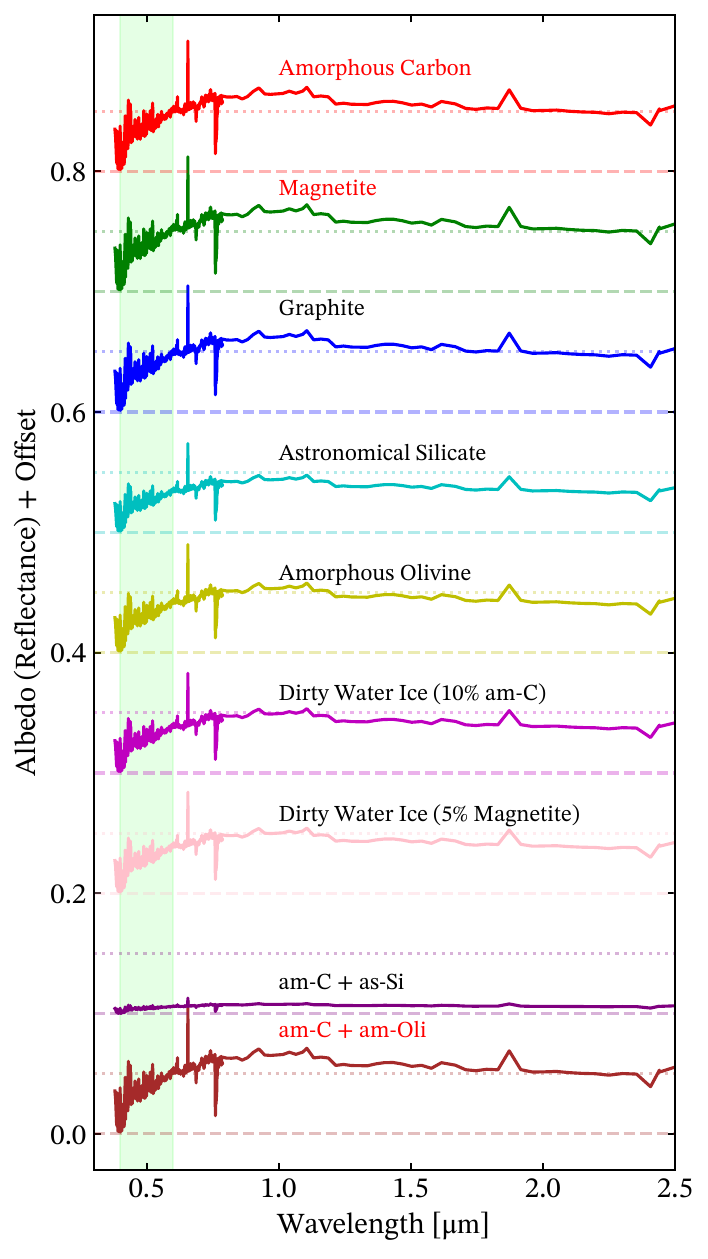}
\caption{
This figure shows the albedo spectra calculated for each dust model, vertically offset for clarity. The dotted and dashed curves show the reflectance equal to 0 and 5\% with the corresponding offsets, respectively. The light-green shaded region marks the wavelength range of the $V$-band. The reliable models (amorphous carbon, magnetite, amorphous carbon (aC)--amorphous olivine (aOli)) are emphasized in red in the annotations.
}
\label{fig:C2}
\end{figure}

Figure~\ref{fig:C2} illustrates the calculated effective reflectance spectra ($p(\lambda)$) based on the dust size distributions derived from the thermal fitting. The effective V-band albedos were derived by averaging these spectra over the wavelength range $0.4$--$0.6~\mu\mathrm{m}$.

\end{document}